\documentclass[10pt,superscriptaddress,twocolumn,amsmath,amssymb,aps,prb,reprint]{revtex4-1}
\usepackage{mathrsfs}
\usepackage{graphicx}
\usepackage{dcolumn}
\usepackage{bm}
\usepackage{color}

\usepackage{epstopdf}

\usepackage[colorlinks,bookmarks=false,citecolor=blue,linkcolor=red,urlcolor=blue]{hyperref}

\newcommand{\be}{\begin{equation}}
\newcommand{\ee}{\end{equation}}
\newcommand{\bea}{\begin{eqnarray}}
\newcommand{\eea}{\end{eqnarray}}

\begin{document}
\title{Enhanced nematicity emerging from higher-order van Hove singularities}
\author{Xinloong Han}\email{hanxinloong@gmail.com}
\affiliation{Kavli Institute for Theoretical Sciences, University of Chinese Academy of Sciences, Beijing 100190, China}
\author{Andreas P. Schnyder}\email{a.schnyder@fkf.mpg.de}
\affiliation{Max-Planck-Institut f\"ur Festk\"orperforschung, Heisenbergstrasse 1, D-70569 Stuttgart, Germany}
 \author{Xianxin Wu}\email{xxwu@itp.ac.cn}
\affiliation{CAS Key Laboratory of Theoretical Physics, Institute of Theoretical Physics, Chinese Academy of Sciences, Beijing 100190, China}

\date{\today}
\begin{abstract}

Motivated by the experimental identification of a higher-order van Hove singularity (VHS) in AV$_3$Sb$_5$ kagome metals, we study electronic instabilities of 2D lattice models with higher-order VHS and flavor degeneracy. In contrast to conventional VHSs, the larger power-law density of states and the weaker nesting propensity of higher-order VHSs conspire together to generate distinct competing instabilities. After discussing the occurrence of higher-order VHSs on square and kagome lattice models, we perform unbiased renormalization group calculations to study competing instabilities and find a rich phase diagram containing ferromagnetism, anti-ferromagnetism, superconductivity and Pomeranchuk orders. Remarkably, there is a generic transition from superconductivity to a $d$-wave Pomeranchuk order with increasing flavor number. Implications for the intriguing quantum states of AV$_3$Sb$_5$ kagome metals are also discussed.

\end{abstract}
 \maketitle

\section{Introduction}
Competing correlated electronic states are a central topic in condensed matter physics. A typical example is the close competition between spin density wave state and $d$-wave superconductivity near half filling in the Hubbard model on the square lattice~\cite{Schulz1987,Dzyaloshinskii1987,Furukawa1988,LeHur2009}. This could provide a possible explanation for the phase diagram of curpate high-T$_c$ superconductors. Importantly, band structures on the square lattice have a saddle point at which the Fermi surface topology changes from hole type to electron type. These saddle points are called van Hove singularity (VHS) points\cite{vanHove1953} and display divergent density of states (DOS) in two dimension (2D). When the Fermi sea is filled up to the VHS, the Fermi surface takes the shape of a parallelogram and good Fermi surface nesting and the divergent DOS generate strong instabilities (double logarithmic) in both particle-particle and particle-hole channels. Similar physics also occur
on the hexagonal lattices with repulsive interactions, where chiral ($d+id$)-wave superconductivity is found to be the ground state at van Hove filling in the formalism of renormalization group~\cite{Chubukov2012,Thomale2012,DHLee2012,Honerkamp}. Recently, another type of VHS, namely the higher-order VHS (HOVHS), was investigated in ABC-stacked trilayer graphene on hexagonal boron nitride\cite{Shtyk2017} and twisted bilayer graphene\cite{LiangFu2019,Isobe2019,Guerci2021}. It features flatter band dispersion along certain directions, while the Fermi surface touch tangentially at the saddle points\cite{Isobe2019}. In the 2D case, the higher-order VHS exhibits power-law divergent DOS, much stronger than the logarithmic one of conventional VHS. Generally, such VHSs are characterized by relatively poor Fermi surface nesting when they are located at time-reversal invariant points. However, the strongly divergent DOS can promote exotic correlated phenomena such as ferromagnetism and supermetal\cite{Isobe2019,Classen2020,Lin2020PRB}.

The corresponding quantum materials convey VHSs play an essential role in revealing the origin of correlated electronic states. Despite tremendous experimental efforts, superconductivity has not been achieved in graphene near VHS filling possibly due to disorder effect. Recently, the new family of kagome metals AV$_3$Sb$_5$ (A=K, Rb, Cs) with V kagome nets\cite{Ortiz2019,Ortiz2020,Neupert,KJiang} has been discovered, which exhibits a higher-order VHS and several conventional VHSs close to the Fermi level. 
The higher-order VHS is located just below the Fermi level and exhibits a flat dispersion along the M-K direction with a dominant quartic term, as confirmed by ARPES experiments\cite{Kang2022,Hu2021}. This multitude of VHSs may be closely related to a number of exotic correlated phenomena that occur in AV$_3$Sb$_5$, such as charge density wave (CDW)\cite{YXJiang2020,ZLiang,Cong2021,Kang2022NM,YongHu2022,HXTan2021,Feng2021,Denner2021,Lin2021,Park2021}, unconventional superconductivity\cite{Zhao2021,Wang2020,Gao2021,Wu2021},
and nematic orders\cite{Gao2021,Zhaoh2021,chenxianhui2022,Li2022,HaiHuWen}. The unconventional nature of superconductivity in AV$_3$Sb$_5$ is evidenced by a large residual thermal conductivity\cite{Zhao2021}, and proximity-induced edge supercurrents\cite{Wang2020} and double superconducting domes under pressure\cite{Chenky2021,Chen2021,Zhang2021}. Below the transition temperature T$_{CDW} = 78-103 $K\cite{YXJiang2020,HXTan2021,ZLiang,Cong2021} AV$_3$Sb$_5$ exhibits an unconventional CDW order with a $2\times 2$ in-plane pattern and time-reversal symmetry breaking, while below T$_{nem} = 35$ K a nematic order emerges~\cite{Gao2021,Zhaoh2021,chenxianhui2022,Li2022,HaiHuWen}. There also exist the sublattice feartures in the kagome metal which relates to the "sublattice interference effect"~\cite{PhysRevB.86.121105} due to the completely different wave function support with filling. Two-fold VHS which have two opposite concavity originated from a pure and a mixed sublattice states near Fermi level, has suggested to have an competition between chiral excitonic state and CDW order~\cite{scammell2023chiral}. Recently, a renormalization group study~\cite{wu2022sublattice} also explored this sublattice feature and showed its important influence on the ordering tendencies.

Higher-order VHSs have also been observed in other materials, such as Sr$_3$Ru$_2$O$_7$\cite{Efremov2019}, $\beta$-YbAlB$_4$~\cite{Ramires2012} and germanene on MoS$_2$\cite{Sante2019}. In addition, twisted bilayer graphene and multi-orbital systems can host VHSs with multiple fermion flavours $N_f$ originating either from valley or orbital degrees of freedom. The increased flavor number can have a significant effect on instabilities, for example, an imaginary CDW order instead of chiral superconductivity is favored on the hexagonal lattices with conventional VHSs for $N_f\geq4$\cite{Lin2019}.

Motivated by the above observations, we systematically study the competing instabilities on the 2D square and hexagonal lattices at higher-order van Hove filling with an SU($N_f$) flavor degeneracy and focus on the effect of the number of flavors. First, based on tight-binding models, we determine the conditions for the occurrence of higher-order VHSs on the square and kagome lattices. With this information we derive general effective models describing the $N_p$ higher-order van Hove saddle points with $N_f$ fermion flavors. Based upon the low-energy models we perform unbiased renormalization group calculations to deduce the competing instabilities. Remarkably, we find that increasing flavor number can drive a generic phase transition from a superconducting order to a nematic order. Finally, we perform mean-field analyses to determine the ground states and discuss the competing orders in the kagome lattice model. Implications for the the AV$_3$Sb$_5$ kagome metals are also discussed.

\section{Higher-order VHS on the square and kagome lattices}
In this section, we derive the conditions for the occurrence of higher-order VHSs in tight-binding models on the square and kagome lattices. This provides a starting point for the renormalization group study in the next section.

\subsection{Square lattice}
We start by discussing VHSs in the square lattice with the dispersion
\begin{eqnarray}
\epsilon_{\bm{k}}&=&-2t(\cos k_x+\cos k_y)-4t'\cos k_x \cos k_y\nonumber\\
&&-2t''(\cos 2k_x+\cos 2k_y)-\mu,
\end{eqnarray}
where $t$, $t'$ and $t''$ are the nearest-neighbor (NN), second NN and third NN hopping amplitudes, respectively. Near the saddle point at X$=(\pi,0)$, the effective dispersion is
\begin{eqnarray}
\epsilon_{\text{X}}({\bf{k}})&=&(-t-2t'+4t'')k^2_x+(t-2t'+4t'')k^2_y\nonumber\\
&&+\frac{t+2t'-16t''}{12}k^4_x+\frac{-t+2t'-16t''}{12}k^4_y \nonumber \\
&&+t'k^2_xk^2_y+4(t'-t'')-\mu.
\label{square1}
\end{eqnarray}
where  \begin{figure}[ht]
\centerline{\includegraphics[width=0.48\textwidth]{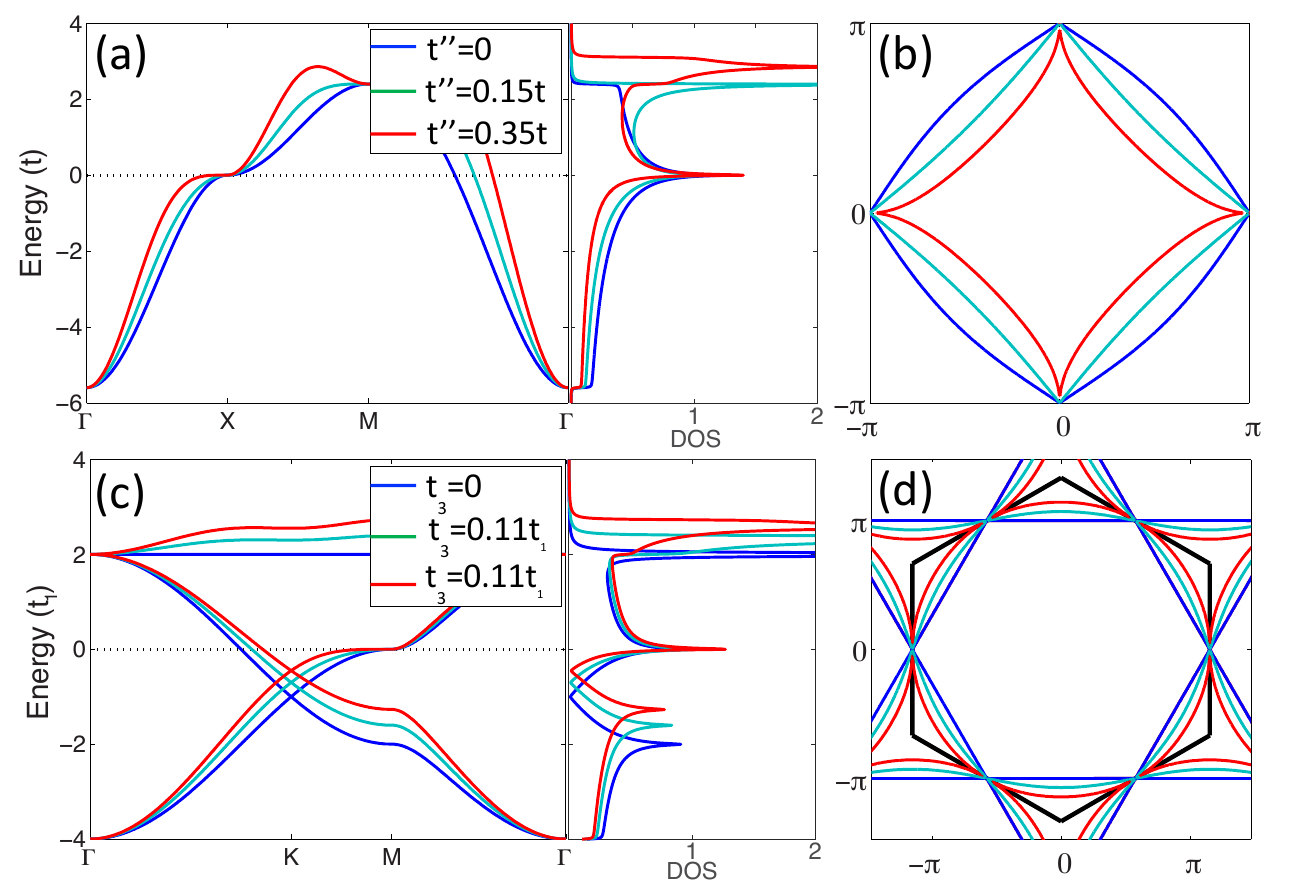}}
\caption{ Band dispersion, density of states (DOS) and Fermi surfaces with variation of hopping parameters for the square lattice (top) and kagome lattice (bottom). A fixed $t'=0.2t$ is adopted with $\mu=4(t'-t'')$ for the square lattice model, while for the kagome lattice model we set $t_2=0$ and $\mu=2t_3$.
 \label{HOVHS} }
\end{figure}$k_{x,y}$ are the momentum relative to the X point. The effective dispersion around the other saddle point Y$=(0,\pi)$ can be obtained by a four-fold rotation and is $\epsilon_{\text{Y}}(\bm{k})=\epsilon_{\text{X}}(k_y,-k_x)$. With only NN hopping, $\epsilon_{\text{X}}(\bm{k})=-t(k^2_x-k^2_y)$ and there are two conventional VHS points at X and Y. When one coefficient of the quadratic terms vanishes in Eq. (\ref{square1}), a conventional VHS point transforms into an higher-order VHS point, i.e. $t''_c=-(\pm t-2t')/4$ with $t'\neq0$. In this case, the dispersion along $k_x$ or $k_y$ become quartic and exhibit a nearly flat dispersion near the VHS point. Fig. \ref{HOVHS} (a) shows the band evolution with a variation of $t''$ for $t'=0.2t$ and $\mu=4(t'-t'')$. It is apparent that the the dispersion along $k_x$ near the X point becomes flatter with increasing $t''$. The corresponding evolution of the Fermi surface at the van Hove filling is displayed in Fig. \ref{HOVHS} (b). At the critical parameter $t''_c=0.35t$, the Fermi surface becomes circular-like centered at $(\pi,\pi)$ and touches tangentially at X and Y points, which is a typical character of higher-order VHS. In the square lattice, there are two higher-order VHSs.

\subsection{ Kagome lattice }
Next, we study the VHSs in kagome lattice models by use of the tight-binding Hamiltonian,
\begin{eqnarray}
\mathcal{H}_0&=&[t_1\sum_{\langle ij\rangle\sigma}c^{\dag}_{i\sigma}c_{j\sigma}+t_2\sum_{\langle\langle ij\rangle\rangle\sigma}c^{\dag}_{i\sigma}c_{j\sigma}+t_3\sum_{\langle\langle\langle ij\rangle\rangle\rangle\sigma}c^{\dag}_{i\sigma}c_{j\sigma}\nonumber\\
&&+h.c.]-\mu\sum_{i\sigma}c^{\dag}_{i\sigma}c_{i\sigma},
\end{eqnarray}
where $t_1$, $t_2$ and $t_3$ denotes the NN, second NN and third NN hopping amplitudes, respectively. In momentum space, the Hamiltonian can be written as $\mathcal{H}_{0}=\sum_{\textbf{k}\sigma}\psi^\dag_{\textbf{k}\sigma}(h(\textbf{k})-\mu)\psi_{\textbf{k}\sigma}$ with $\psi^\dag_{\textbf{k}\sigma}=[c^\dag_{A\sigma}(\textbf{k}),c^\dag_{B\sigma}(\textbf{k}),c^\dag_{C\sigma}(\textbf{k})]$ and $A,B,C$ being the three sublattice indices. The Hamiltonian matrix elements are,
\begin{eqnarray}
h_{11}&=&2t_3\cos k_y,\nonumber\\
h_{22/33}&=&2t_3\cos(\frac{\sqrt{3}}{2}k_x\pm\frac{1}{2}k_y),\nonumber\\
h_{12/13}&=&2t_1\cos(\frac{\sqrt{3}}{4}k_x\mp\frac{1}{4}k_y)+2t_2\cos(\frac{\sqrt{3}}{4}k_x\pm\frac{3}{4}k_y),\nonumber\\
h_{23}&=&2t_1 \cos\frac{1}{2}k_y+2t_2cos\frac{\sqrt{3}}{2}k_x,
\end{eqnarray}
with $h_{\beta\alpha}=h^*_{\alpha\beta}$ for $\alpha\neq\beta$.
The band dispersion features a flat band and two Dirac cones. Moreover, there are two types of VHSs located at $E=2t_3$ and $E=2(t_1-t_2-t_3)$, featuring pure sublattice (p-type) or mixed sublattice (m-type) characters, respectively~\cite{Wu2021}. In order to explore the realization of higher-order VHSs on kagome lattice models, we derive a low-energy effective model near the saddle points. The effective Hamiltonian around the M$_1=(\frac{2\pi}{\sqrt{3}},0)$  point reads,
\begin{widetext}
 \begin{eqnarray*}
 h_{M_1}(\bm{q})&=& \left(\begin{array}{ccc}
2t_3(1-\frac{1}{2}q^2_y) & cq_x+dq_y & cq_x-dq_y  \\
cq_x+dq_y& -2t_3[1-\frac{1}{2}(\frac{\sqrt{3}}{2}q_x+\frac{1}{2}q_y)^2]   & 2(t_1-t_2)-\frac{t_1}{4}q^2_y+\frac{3t_2}{4}q^2_x \\
cq_x-dq_y & 2(t_1-t_2)-\frac{t_1}{4}q^2_y+\frac{3t_2}{4}q^2_x & -2t_3[1-\frac{1}{2}(\frac{\sqrt{3}}{2}q_x-\frac{1}{2}q_y)^2] \\
\end{array}\right),
 \end{eqnarray*}
 \end{widetext}
where $c=-\sqrt{3}(t_1+t_2)/2$, $d=(t_1-3t_2)/2$ and $\bm{q}=(q_x,q_y)$ is the momenta relative to the M$_1$ point. For the p-type VHS with $E=2t_3$, the energy dispersion up to the quadratic order is,
\begin{eqnarray}\label{Eq:eq5}
E^{M_1}_p(\bm{q})&=&2t_3-\frac{3(t_1+t_2)^2}{4(t_1-t_2-2t_3)}q^2_x\nonumber\\
&+&\big[\frac{(t_1-3t_2)^2}{4(t_1-t_2+2t_3)}-t_3\big]q^2_y.
\end{eqnarray}
The effective dispersion around $M_2=(\frac{\pi}{\sqrt{3}},\pi)$ and $M_3=(-\frac{\pi}{\sqrt{3}},\pi)$ can be obtained by six-fold rotations, namely $E^{M_2}_p(\bm{q})=E^{M_1}_p(\hat{C}_6\bm{q})$ and $E^{M_3}_p(\bm{q})=E^{M_1}_p(\hat{C}_3\bm{q})$.
With only nearest-neighbor hopping, i.e. $t_2=t_3=0$, this VHS is conventional and features perfect Fermi surface nesting with $E^{M_1}_p=2t_3-\frac{t_1}{4}(3q^2_x-q^2_y)$. In order to realize a higher-order VHS, that is, one of two quadratic terms in Eq. (\ref{Eq:eq5}) must vanish. As the quadratic and quartic terms of $q_x$ vanish simultaneously with $t_1=-t_2$, we consider the $q_y$ terms. Setting the coefficient of $q^2_y$ to zero, we obtain $t_{3c}=\frac{1}{4} \left(-t_1+t_2\pm\sqrt{3 t_1^2-14 t_2 t_1+19 t_2^2}\right) $. The corresponding evolution of band structures and Fermi surfaces at the VHS filling is displayed in Fig. \ref{HOVHS} (c) and (d). At the critical parameter $t_{3c}$, the band dispersion along M-K is quartic and the corresponding Fermi surfaces are circular-like around the center. In particular, for $t_3=0$, the coefficients for $q^2_y$ and $q^4_y$ will vanish simultaneously, generating a higher-order VHS with more strongly divergent DOS. The DOS for a generalized dispersion $\epsilon(\bm{q})=aq^2_x-bq^{2\alpha^{\prime}}_y+...$ ($\alpha^{\prime}>1$) is\cite{Isobe2019,Classen2020},
\begin{eqnarray}
\rho(\epsilon)=\left\{
\begin{aligned}
\rho_+\epsilon^{-\kappa}, \text{for }\epsilon>0\\
\rho_-\epsilon^{-\kappa}, \text{for }\epsilon<0\\
\end{aligned}
\right
.
\end{eqnarray}
where $\kappa=1/2-1/(2\alpha^{\prime})$, $\rho_+=\Gamma[1/(2\alpha^{\prime})]\Gamma[1/2-1/(2\alpha^{\prime})]/(4\alpha^{\prime}\pi^{5/2}a^{1/2}b^{1/{2\alpha^{\prime}}})$ and $\rho_-=\rho_+\sin[\pi/(2\alpha^{\prime})]$. The DOS is asymmetrical with respect to zero, as shown in Fig. \ref{HOVHS} (a) and (c) from numerical calculations. When only the quadratic term vanishes, $\alpha^{\prime}=2$ and $\kappa=1/4$. While both quadratic and quartic terms vanish, $\alpha^{\prime}=3$ and $\kappa=1/3$.

For the m-type VHS with $E=2(t_1-t_2-t_3)$, the energy dispersion up to the second order at M$_1$ point is,
\begin{eqnarray}
E^{M_1}_m(\bm{q})&=&2(t_1-t_2-t_3)+\frac{3(t_1-t_3)(t_1+3 t_2+2 t_3)}{4(t_1-t_2-2t_3)}q^2_x\nonumber\\
&+&\frac{1}{4}(t_3-t_1)q^2_y.
\end{eqnarray}
Within this model, the vanishing of the quadratic term will lead to a completely flat dispersion along the corresponding direction. A higher-order VHS can be realized upon the inclusion of longer-range hopping. On the kagome lattice, there are three inequivalent M points thus three saddle points.

\section{The general Model}\label{Model}

We further investigate the competing instabilities at the higher-order van Hove filling. Since the unbiased RG method takes into account the particle-particle and particle-hole channels on equal footing,  and the low-energy physics is dominated by the electron scattering in the vicinity of saddle points, we simplify the above models into patch models, which only consider the electrons near the saddle points\cite{Furukawa1988,Chubukov2012,Yao2015}, and then perform a RG analysis. The general low-energy theory with total $N_p$ patches and $N_f$ flavors of fermions with the SU($N_f$) symmetry is given by\cite{Chubukov2012,Classen2020},
\bea
\mathcal{L}=&&\sum_{\alpha\sigma}\psi_{\alpha\sigma}^{\dagger}(\partial_{\tau}-\epsilon_{\alpha,{\bf k}}+\mu)\psi_{\alpha\sigma}+\sum_{\sigma\sigma^{\prime}}\Big\{\sum_{\alpha}g_4\psi^{\dagger}_{\alpha\sigma}\psi^{\dagger}_{\alpha\sigma^{\prime}}\nonumber \\
&&\times\psi_{\alpha\sigma^{\prime}}\psi_{\alpha\sigma}+\sum_{\alpha\neq \alpha^{\prime}}\big[g_1\psi_{\alpha\sigma}^{\dagger}\psi^{\dagger}_{\alpha^{\prime}\sigma^{\prime}}\psi_{\alpha\sigma^{\prime}}\psi_{\alpha^{\prime}\sigma}+g_2\psi_{\alpha\sigma}^{\dagger}\nonumber \\
&&\times\psi^{\dagger}_{\alpha^{\prime}\sigma^{\prime}}\psi_{\alpha^{\prime}\sigma^{\prime}}\psi_{\alpha\sigma}+g_3\psi_{\alpha\sigma}^{\dagger}\psi^{\dagger}_{\alpha\sigma^{\prime}}\psi_{\alpha^{\prime}\sigma^{\prime}}\psi_{\alpha^{\prime}\sigma}\big]\Big\},
\eea
where\begin{figure}[ht]
\includegraphics[width=0.48\textwidth]{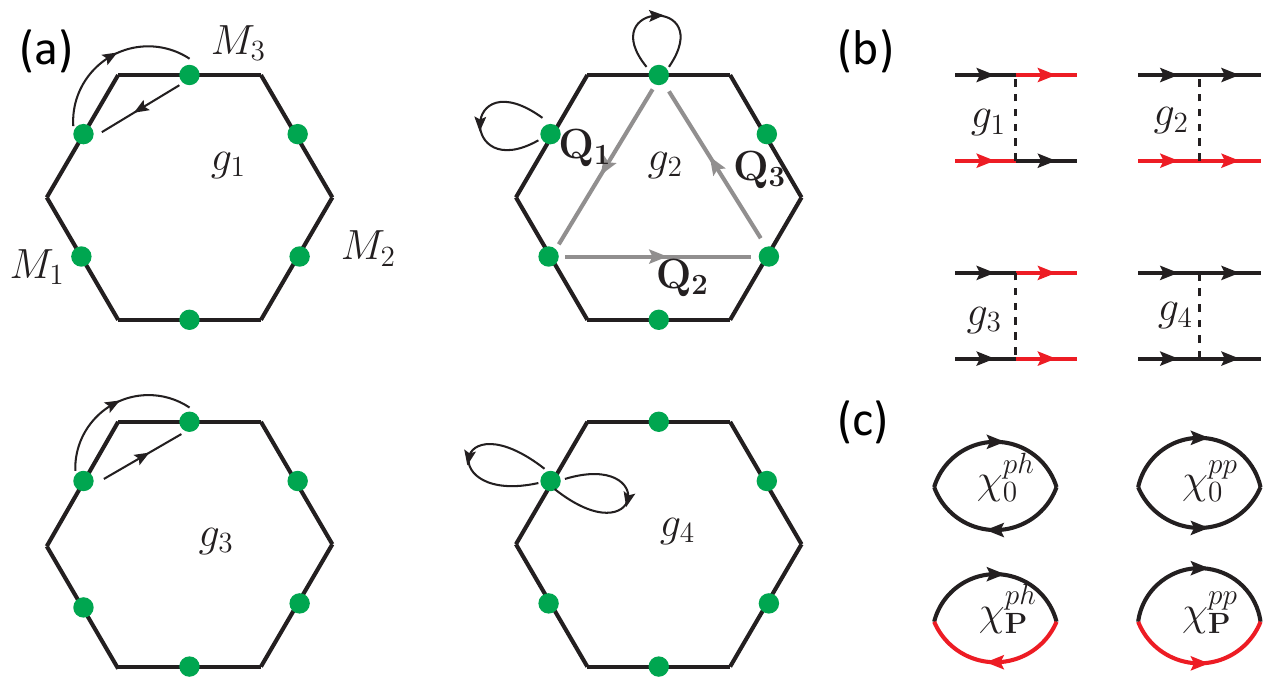}
\caption{(a) Four different scattering processes in the kagome lattice with three different patches. Here ${\bf Q}_1$, ${\bf Q}_2$ and ${\bf Q}_3$ represent the nesting vectors between two patches. (b) Feynman diagrams describing these four scattering processes. The lines with different colors represent fermions in different patches. (c)
Feynmann diagrams representing the particle-hole and particle-particle polarizations with zero or finite momentum transfer ${\bf P}={\bf Q}_{1,2,3}$ at the zero order.}
\label{fig:interactions}
\end{figure} the patch index $\alpha,\alpha^{\prime}=1,\dots,N_p$, and the spin/flavor index $\sigma,\sigma^{\prime}=1,\dots,N_f$. $\psi_{\alpha\sigma}$ ($\psi^{\dagger}_{\alpha\sigma}$) are electron annihilation (creation) operators with spin/flavor $\sigma$ at patch $\alpha$. Here $\epsilon_{\alpha,{\bf k}}$ is the energy dispersion of electrons expanded up to $2n$-th order at patch $\alpha$, where $n$ is a positive integer ($n=1$ for conventional VHS). The chemical potential is denoted by $\mu$. In the rest of the paper, we take $\mu=0$ and $n=2$. The short-range interactions $g_i$ which describe different scattering processes are illustrated in the Fig. \ref{fig:interactions}. One should note the Umklapp scattering $g_4$ appear when the nesting momentum $2{\bf Q}=0$ up to a reciprocal lattice vectors. The inclusion of only short-range interactions in the low-energy description is justified because the highly metallic screening close to VHSs suppresses long-ranged Coulomb interactions.

The higher-order VHSs possess distinct properties compared with conventional VHSs. First, the corresponding DOS at a given patch diverges with a power law, i.e., faster than the logarithmic divergence at a conventional VHS. And second, the nesting between different patches is weaker compared to the case with conventional VHSs. These two physcial properties can be checked from the zero momentum and finite momentum (difference between distict patches) 
\begin{figure}[ht]
\includegraphics[width=0.48\textwidth]{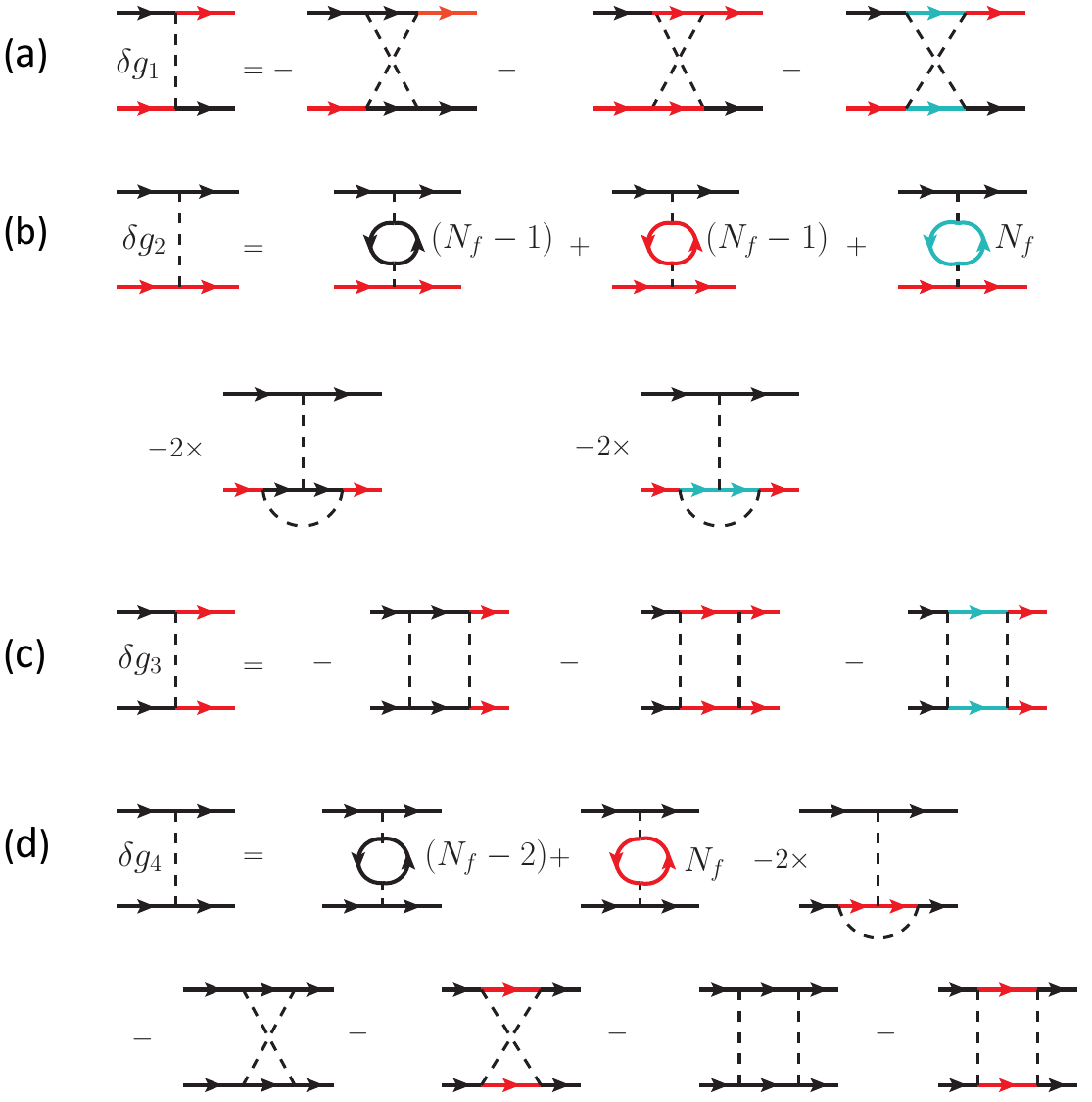}
\caption{Feynman diagrams for the one-loop corrections of the four different scattering processes $g_i$ with the inclusion of dominant susceptibilities. The lines with different colors denote the different patches.  (a) To the leading order, only the crossed diagrams contribute to the backward scattering process between two patches. (b) Two different topological Feynman diagrams contribute to the renormalization of the inter-patch forward scattering process $g_2$ between two different patches. (c) There are three BCS type diagrams to correct the umklapp scattering $g_3$. (d) The diagrams to renormalize the intra-patch forward scattering $g_4$.}
\label{fig:FeynmanDiagrams}
\end{figure}susceptibilities. The zero momentum particle-particle and particle-hole susceptibilities can be defined as $\chi_{0}^{ph}(\Lambda/T)=-T\sum_{i\omega_m}\int \frac{d^2 {\bf k}}{(2\pi)^2}G^{0}_{\alpha}(i\omega_m,{\bf k}) G^{0}_{\alpha}(i\omega_m,{\bf k})$ and $\chi_{0}^{pp}(\Lambda/T)=T\sum_{i\omega_m}\int \frac{d^2 {\bf k}}{(2\pi)^2}G^{0}_{\alpha}(-i\omega_m,-{\bf k}) G^{0}_{\alpha}(i\omega_m,{\bf k})$, respectively. Note we have hided the patch index $\alpha$ in susceptibility $\chi^{ph,pp}_0$ because of the lattice symmetry. Here $G_{\alpha}^{0}(i\omega_m,{\bf k})=1/(i\omega_m-\xi_{\alpha,{\bf k}})$ is the bare Green's function of electrons at patch $\alpha$ with $\xi_{\alpha,{\bf k}}=\epsilon_{\alpha,{\bf k}}-\mu$. $T$ is the temperature of the system, and the magnitude of momentum has an ultraviolet  cutoff $\Lambda$. Matsubara-frequency for the electrons is taken as $i\omega_m=i(2m+1)T$ with $m=0,\pm 1,\pm 2,\dots,$. Focusing on the van Hove filling with $\mu=0$, we find that susceptibilities with zero momentum transfer exhibit the same power-law divergent behavior but with different prefactors, i.e., with $\chi_{0}^{pp},\chi_{0}^{ph}\propto(\rho_++\rho_-)/(2T^{\kappa})$. While the finite momentum susceptibilities $\chi_{{\bf P}_{\alpha\beta}}^{ph}(\Lambda/T)=-T\sum_{i\omega_m}\int \frac{d^2 {\bf k}}{(2\pi)^2}G^{0}_{\alpha}(i\omega_m,{\bf k}) G^{0}_{\beta}(i\omega_m,{\bf P}_{\alpha\beta}+{\bf k})$ and $\chi_{{\bf P}_{\alpha\beta}}^{pp}(\Lambda/T)=T\sum_{i\omega_m}\int \frac{d^2 {\bf k}}{(2\pi)^2}G^{0}_{\alpha}(-i\omega_m,-{\bf k}) G^{0}_{\beta}(i\omega_m,{\bf P}_{\alpha\beta}+{\bf k})$ where patches $\alpha\neq \beta$ and finite momentum ${\bf P}_{\alpha\beta}={\bf Q}_{1,2,3}$ as shown in Fig. \ref{fig:interactions} (a),  exhibit quite different behavior, $\chi_{{\bf P}_{\alpha\beta}}^{ph} \propto$ const, and $\chi_{{\bf P}_{\alpha\beta}}^{pp}\propto \ln(\Lambda/T)$~\cite{Classen2020}. Therefore, approaching zero temperature, the zero momentum susceptibility $\chi_0^{pp/ph}$ diverges much faster than the ones with finite momentum $\chi_{\bf P}^{pp/ph}$, in contrast to conventional VHSs. These characteristics play an essential role in determining the competing orders. Note we have hided the patch index $\alpha$ in susceptibility $\chi^{ph,pp}_0$ because of the lattice symmetry.

\section{RG flow equations}
After presenting the effective model and analysis of susceptibilities, we turn to study the competing instabilities by performing renormalization group calculations. Within the renormalization group (RG) approach, high-energy degrees of freedom are iteratively integrated out, such that the initial microscopic description gradually evolves to an effective low-energy model with decreasing energy cutoff. The initial microscopic description and the effective low-energy model have the same form, except for the coupling constants which depend on a running parameter, i.e., the RG flow time. The coupling constants can grow or shrink under the RG flow, which are relevant and irrelevant couplings, respectively. Approaching a critical RG time, one or several coupling constants can diverge and the system goes to a strong-coupling regime. Usually the susceptibility of certain order will also diverge, indicating that the system undergoes a phase transition to an ordered state. In fermionic systems interactions with finite frequency and/or momentum away from the Fermi surface are irrelevant, such that only zero-frequency coupling constants at the Fermi surface need to be considered~\cite{Metzner2012,Qin2019}.

For these higher-order VHSs, the corrections of the coupling constants from both tree-level and one-loop terms are important and dimensionless coupling constants $\tilde{g}_i=g_i \partial \chi^{pp}_0/\partial s$ can be introduced with $s=\ln{(\Lambda/T)}$. Generally, there are no stable fixed points and thus we focus on the strong-coupling behavior of the dimensional couplings $g_i$~\cite{Classen2020,Lin2020PRB}, where the tree-level terms are irrelevant.
The RG flow equations with the one-loop terms can be obtained directly by implementing the Feynman diagrams as shown in Fig. \ref{fig:interactions} with the inclusion of dominant susceptibilities in the particle-particle and particle-hole channels. Here, we derive general RG flow equations for $N_p$ patches and $N_f$ flavors, such that different lattice structures and different numbers of orbital and spin degeneracies can be captured. The same result is also derived in Ref. \cite{Classen2020}. As depicted by Fig. \ref{fig:interactions}, the fermions in different patches are denoted by different colors, then we have to sum over the remaining fermion patch numbers. Since there is no flavor flip process, the summation over the fermion flavor only occur in the fermion loop. To simplify the identification of the critical temperature, we adopt $y=\chi_{0}^{pp}(T)-\chi_{0}^{pp}(\Lambda)=\rho_0 f(\kappa)\Lambda^{-\kappa}((\Lambda/T)^{\kappa}-1)$ as the RG flow time, with $\rho_0=(\rho_+ \rho_-)/2$ and $f(\kappa)=\frac{1}{2}\int d\epsilon |\epsilon|^{-(1+\kappa)}\text{tanh}(\epsilon/2) $~\cite{Classen2020}. Straightforwardly, we obtain the RG flow equations after considering the one-loop corrections in Fig. \ref{fig:FeynmanDiagrams},
\bea
\frac{d g_1}{dy}=&&d_0g_1\big[2 g_4+(N_p-2) g_1\big],\label{RGeq1} \\
\frac{d g_2}{dy}=&&-d_0(N_p-2) g_2\big[N_fg_2-2g_1\big]\nonumber \\
&&-2d_0g_4\big[(N_f-1) g_2-g_1\big],\label{RGeq2}\\
\frac{d g_3}{dy}=&&-g_3\big[(N_p-2)g_3+2g_4\big], \label{RGeq3}\\
\frac{d g_4}{dy}=&&-d_0\Big\{ (N_p-1)\big[N_f g_2^2-2g_1g_2 -g_1^2\big]  \nonumber \\
&&+(N_f-3) g_4^2\Big\}-g_4^2-(N_p-1)g_3^2. \label{RGeq4}
\eea
Here, the nesting parameter $d_0 = \partial\chi_0^{ph}/\partial\chi_0^{pp}$. It is noted that we have ignored subdominant terms related to susceptibilities with large momentum transfer, as they have weaker divergence than $\chi_0^{pp/ph}$. The RG equations have two prominent features: (1) the RG equations for $g_1$ and $g_3$ contain $g_1$ and $g_3$, respectively; (2) the RG functions for $g_1$ and $g_2$ do not contain $g_3$ and that of $g_3$ does not contain $g_1$ and $g_2$, in contrast to the RG equations of conventional VHSs~\cite{Chubukov2012}. The first feature is related to a symmetry: all interactions except $g_1$ conserve spin at each saddle point separately and all interaction except $g_3$ conserve the number of electrons at each saddle point separately. Consequently, the RG equation of a nonconversing interactions $g_{1,3}$ must contain a factor of itself, indicating that the sign of $g_1$ and $g_3$ will be fixed during the RG flow. The second feature is related to the neglect of subdominant terms and suggests that $g_1$ and $g_2$ are fully decoupled with $g_3$ during the RG evolution.

The magnitude of $d_0$ dramatically affects the flow of the RG, determining which coupling constant will diverge first. One can check this by considering the limit of small $d_0 \to 0$ and the limit of large $d_0 \to \infty$, respectively. When $d_0=0$, the coupling constants $g_1$ and $g_2$ will not flow, staying at the given initial values for the RG time $y$ running from the ultraviolet (UV) value to the infrared (IR) value. The negative definite character of the differential equation Eq. (\ref{RGeq4}) for $g_4$ at $d_0=0$ will ensure that the coupling constant $g_4$ always flows to negative infinity, except for the trivial fixed point $(g_3, g_4) = (0, 0)$. Thus, in the limit of small $d_0$ superconducting order is favored, while the instability in the particle-hole channel, such as charge-density wave (CDW), ferromagnetic (FM) and anti-ferromagnetic (AFM) order, is completely suppressed. In contrast, in the large $d_0$ limit the  particle-hole instabilities are favored, while the particle-particle channel leading to superconductivity is suppressed. The coupling constant $g_i$ that diverges first at the RG time $y_c$ determines the leading order and the corresponding transition temperature is $T_c\propto\big(\rho_0 f(\kappa)/(\rho_0f(\kappa)+\Lambda^{\kappa}y_c) \big)^{1/\kappa}$. We note that this power-law behavior of $T_c$ in systems with HOVHs is significantly different from the exponential behavior of $T_c$ in systems with conventional VHSs~\cite{Furukawa1988,Chubukov2012}.

\section{Ordering instabilities}

To identify the leading instabilities, we introduce the test vertices in various particle-particle and particle-hole channels, 
$\delta \mathcal{L}=\sum[\Delta{\psi}^{\dagger} {\psi}^{(\dagger)}+h.c.]$. In the particle-particle channel, we consider the flavor pairing $\Delta_\alpha[{\psi}^{\dagger}_{\alpha\sigma} {\psi}^{\dagger}_{\alpha\sigma'}-{\psi}^{\dagger}_{\alpha\sigma'} {\psi}^{\dagger}_{\alpha\sigma}]$ with $\sigma>\sigma'$ and it is antisymmetric in the flavor space. In the particle-hole channel, the charge Pomeranchuk order is the summation of uniform flavor terms $\sum_{\sigma}{\psi}^{\dagger}_{\alpha\sigma} {\psi}_{\alpha\sigma}$ and the magnetic order is the summation of the antisymmetic terms $\sum_{\sigma>\sigma'}[{\psi}^{\dagger}_{\alpha\sigma} {\psi}_{\alpha\sigma'}-{\psi}^{\dagger}_{\alpha\sigma'} {\psi}_{\alpha\sigma}]$. Under the RG flow, these test vertices
obtain one-loop corrections through the divergent susceptibilities and the corresponding RG equation reads,
\bea
\frac{d\Delta_{\alpha}(y)}{dy}=V_{\alpha\beta}(y)\Delta_{\beta}(y).
\eea
These orders can be further decoupled into different irreducible channels according to the lattice symmetry, $\dot{\Delta}^I(y)=V_{\Delta^I}(y)\Delta^I(y)$ where $V_{\Delta^I}(y)$ is the interaction vertex for $\Delta^I$ and is a linear combination of coupling constants. Approaching the critical RG time, the interaction $g_i$ usually diverges as $g_i(y)=G_i/(y_c-y)$, hence $V_{\Delta^I}(y)$ diverges as $V_{\Delta^I}(y)=\lambda_I/(y_c-y)$ where $\lambda_I$ is a constant number. Correspondingly, the vertex scales as $\Delta^I(y) \propto (y_c-y)^{-\lambda_{I}}$ and the susceptibility $\chi_I$ related to $\Delta_I$ scales as $\chi_{I}(y)\propto (y_c-y)^{-\eta_{I}}$ with $\eta_{I}=2\lambda_{I}-1$ from the equation $d \chi_{I}/dy \propto |\Delta_{I}|^2$. The parameters for two representative orders are $\lambda_{dSC}=G_3-G_4$ for the d-wave pairing and $\lambda_{dPOM}=-d_0G_1+d_0 N_fG_2-d_0(N_f-1)G_4$ for the d-wave Pomeranchuk order. The order with the most positive $\lambda_{I}$ diverges fastest and is the leading instability. We summarize the interactions and their coefficients for various competing orders in particle-particle and particle-hole channels in the Table. \ref{table:InteractionV}.
\begin{table}
	\centering
	\caption{Competing orders and their divergent coefficients $\lambda_{I}$. The orders in three channels can be $s$-wave and $d$-wave depending on their symmetry and they are labelled as sCO or sCO with CO=POM, FM SC. In the square lattice, these orders are singlet, while in the hexagonal lattice, the $d$-wave orders are doublets. 
	}
	\label{table:InteractionV}
	\begin{tabular}{c|c}
	\hline
	\hline
	Competing oder & coefficient $\lambda_{I}$ \\
	\hline
	sPOM  & $d_0(N_p-1)(G_1-N_fG_2)-d_0(N_f-1)G_4$  \\
	\hline
	dPOM & $d_0(-G_1+N_f G_2) -d_0(N_f-1)G_4$  \\
	\hline
	FM &  $d_0(N_p-1) G_1 +d_0G_4$  \\
	\hline
	dFM & $-d_0G_1+d_0 G_4$ \\
	\hline
	sSC & $-(N_p-1)G_3-G_4$ \\
	\hline
 dSC & $G_3-G_4$ \\
	\hline
	\hline
	\end{tabular}
\end{table}

\section{Fixed trajectories and Stabilities}\label{sec:FixedTandStb}
In the RG equations (\ref{RGeq1})-(\ref{RGeq4}) for the higher-order VHSs, there is no positive definite coupling constant, distinct from the case of conventional VHS~\cite{Chubukov2012}. The RG flow of differential equations (\ref{RGeq1})-(\ref{RGeq4}) exhibit fixed trajectories along which the couplings become singular but their ratios tend to finite values. Along these fixed trajectories the couplings $g_i$ diverge as
$g_i(y)=\frac{G_i}{y_c-y}$,
as mentioned before. To analyze the stability of these fixed trajectories, it is useful to introduce the relative couplings $\gamma_i=g_i/g_4$ with $i=1,2,3$. The reason that we choose $g_4$ as the denominator is that the $g_4$ always diverges except for the trivial unstable initial condition $g_i(0)=0$. This means that $|g_4|$ monotonically grows near critical RG time $y_c$. Thus we choose $x=\ln|g_4(y)|$ as a running parameter and the reduced RG equations for $\gamma_i$ ($i=1,2,3$) for given $N_p=2$ as a typical example are given by
\begin{figure}[ht]
\includegraphics[width=0.45\textwidth]{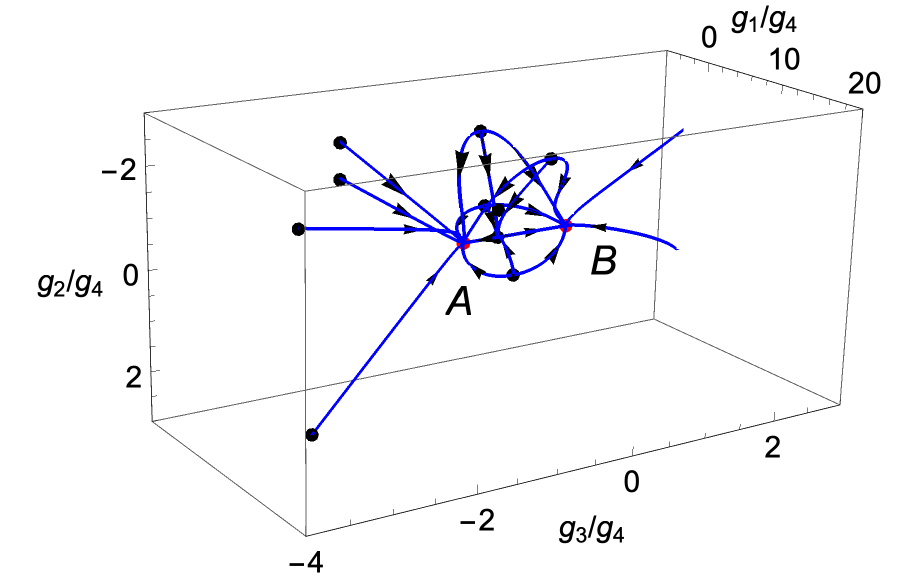}
\caption{$3$D RG flow diagram of ($\frac{g_1}{g_4}$,$\frac{g_2}{g_4}$,$\frac{g_3}{g_4}$) under RG time $y$ with a given initial $g_4(y=0)=-0.1$, number of fermion flavors $N_f=4$ and patches $N_p=3$, and the nesting parameter $d_0=0.1$. Here the black points denote the unstable fixed points for the reduced RG equations, and they eventually flow to the stable points indicated by red points, which are labeled as $A$ and $B$, respectively.}
\label{fig:3Dflows}
\end{figure}
\begin{figure*}[ht]
\includegraphics[width=0.9\textwidth]{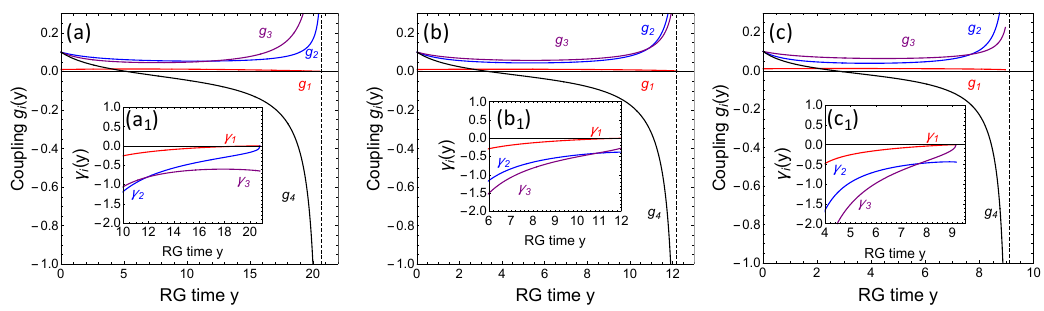}
\caption{RG flows of couplings $g_i(y)$ and reduce parameters $\gamma_i(y)$ with initial condition $g_1(0)=0.01$ and $g_{2}(0)=g_3(0)=g_4(0)=0.1$ for the hexagonal lattice $N_p=3$. Here the nesting parameter is taken as $d_0=0.5$. (a),(b),(c) shows the RG flows of the couplings with RG time y for fermion flavors $N_f=2$, $4$ and $6$, respectively. The black dashed line denote the critical RG time $y_c$ where some coupling strength diverge. The inset display the RG flows of $\gamma_i(y)$ near critical RG time $y_c$. $(\gamma_1(y),\gamma_2(y),\gamma_3(y))$ tend to flow to a stable fixed point at a given initial parameters for the reduced RG equations (\ref{ReducedRGeq1})-(\ref{ReducedRGeq3}).}
\label{fig:RG flows}
\end{figure*}
\bea
&&\beta_1(\{\gamma_i\})=\frac{d \gamma_1}{d x}=\frac{2d_0 \gamma_1}{f(\gamma_1,\gamma_2,\gamma_3)}-\gamma_1, \label{ReducedRGeq1}\\
&&\beta_2(\{\gamma_i\})=\frac{d \gamma_2}{d x }=\frac{-2d_0 [(N_f-1)\gamma_2-\gamma_1]}{f(\gamma_1,\gamma_2,\gamma_3)}-\gamma_2,\label{ReducedRGeq2}\\
&&\beta_3(\{\gamma_i\})=\frac{d \gamma_3}{d x}=\frac{-2\gamma_3}{f(\gamma_1,\gamma_2,\gamma_3)}-\gamma_3, \label{ReducedRGeq3}
\eea
where $f(\gamma_1,\gamma_2,\gamma_3)=-d_0 N_f \gamma_2^2+2d_0\gamma_1\gamma_2+d_0\gamma_1^2-\gamma_3^2-d_0(N_f-1)-1$. Since in this system $G_4$ is always non vanishing (we ignore the trivial solution $G_4=0$), then the fixed points for the reduced RG equations are $(\Gamma_i=G_i/G_4)$. The reason we analyze the reduced RG equations (\ref{ReducedRGeq1}-\ref{ReducedRGeq3}) is that determining the stability conditions of the fixed trajectories in RG equations (\ref{RGeq1}-\ref{RGeq4}) can be obtained by studying the stability condition of its corresponding fixed points in reduced RG equations. The matrix $S_{ij}=(\partial \beta_{i}/\partial \gamma_j)|_{(\Gamma_l)}$ is so-called stability matrix\cite{Herbut2007} and the signs of its eigenvalues determine the stability of the fixed point $(\Gamma_1,\Gamma_2,\Gamma_3)$. If all eigenvalues of the matrix $\bf{S}$ at the fixed point $(\Gamma_i)$ are non-positive, then the fixed point $(\Gamma_i=G_i/G_4)$ or the corresponding trajectory $(G_1,G_2,G_3,G_4)$ is immune to perturbation around the fixed point, in other words, it is a stable solution. For the square lattice with $N_p=2$ patches and general flavor number $N_f$, the analysis of the stability of fixed points is straightforward and we give all fixed points, their stable conditions and corresponding dominant orders in the Appendix \ref{FixedTSquare}. We find that instabilities in both particle-particle and particle-hole channels can be leading but a large $N_f$ favors particle-hole orders. For the hexagonal lattice with $N_p=3$, we can obtain all fixed trajectories provided in the Appendix \ref{appendixB} but the analysis for conditions of stability and competing orders need numerical calculations. For the more general case with fermion patches $N_p$ and flavors $N_f$, the analytical trajectories are difficult to obtain but one can identify the fixed points and their stability from the 3D RG flow diagram through numerical calculations. For example, we plot such a diagram in Fig. \ref{fig:3Dflows} for the case of hexagonal lattice with $N_p=3$ and $N_f=4$ (two fold VHSs), where the initial interactions are $g_4(y=0)=-0.1$ and the nesting parameter is $d_0=0.1$. From Fig. \ref{fig:3Dflows}, it is apparent that the unstable fixed points denoted by black points will flow to one of the fixed points denoted by red point with a small perturbation. The relative interactions $(\Gamma_1,\Gamma_2,\Gamma_3)$ of two stable points, labeled by $A$ and $B$, are $(0,0,-0.466)$ and $(0,0,0.966)$, respectively. The corresponding interactions $(G_1,G_2,G_3,G_4)$ are $(0,0,0.304,-0.652)$ and $(0,0,-0.326,-0.377)$, respectively. Therefore, from the coefficients $\lambda$ of the competing orders listed in the Table. \ref{table:InteractionV}, we find that the dominant instabilities at the fixed points $A$ and $B$ are $d$- and $s$-wave superconducting states, respectively.

\section{Phase transition driven by the number of flavors}

For conventional VHSs, the leading instabilities significantly depend on the number of patches (type of lattice), even for the same initial interaction setting. For example, with initial repulsive interactions, the RG analysis reveals that the d-wave superconducting state and spin density wave (SDW) state are competing on the square lattice\cite{LeHur2009}, while the chiral $d$-wave superconducting states become the dominant on the hexagonal lattice \cite{Chubukov2012}. The number of flavor also has a dramatic effect on the competing states even at conventional van Hove filling~\cite{Lin2019}. In this section, we explore the effect of number of flavors and lattice type competing states for higher-order VHSs through the RG analysis.
\begin{figure}[ht]
\includegraphics[width=0.45\textwidth]{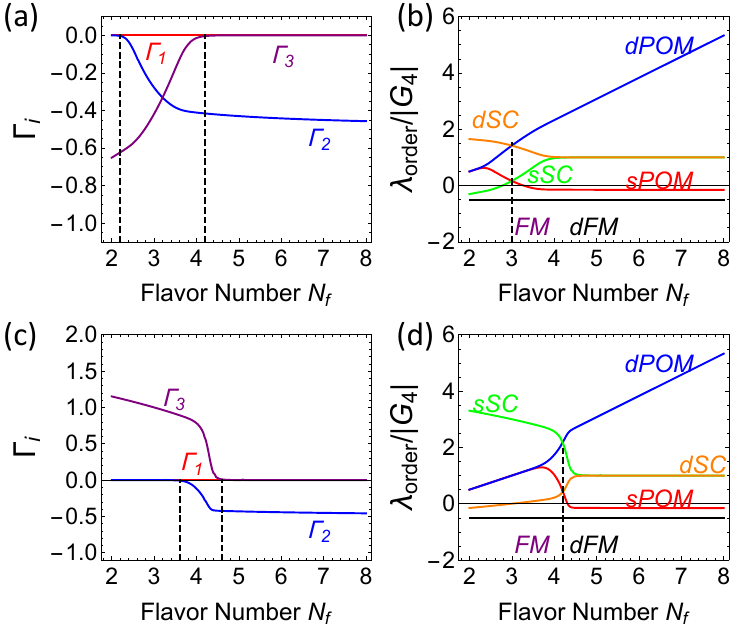}
\caption{On the hexagonal lattice with $N_p=3$, the solution $\Gamma_i=G_i/G_4$ and the ration of susceptibility coefficient $\lambda$ and $|G_4|$ for different competing orders as functions of fermion flavors $N_f$ with fixed nesting parameter $d_0=0.5$. The two black dashed lines separate three regions $(0,0,G_3,G_4)$, $(0,G_2,G_3,G_4)$ and $(0,G_2,0,G_4)$ for the curves for $\Gamma_i$. And the dashed black line in curves $\lambda_{order}/|G_4|$ denotes the position where the phase transition occurs. (a) and (b) are solutions for initial coupling constants $g_{2,3,4}(0)=0.1$ and $g_1(0)=0.01$. (c) and (d) are solutions for $g_{2,4}(0)=0.1$, $g_{1}(0)=0.01$ and $g_3(0)=-0.1$.}
\label{fig:GammaOrder}
\end{figure}
\begin{figure*}[ht]
\includegraphics[width=1.\textwidth]{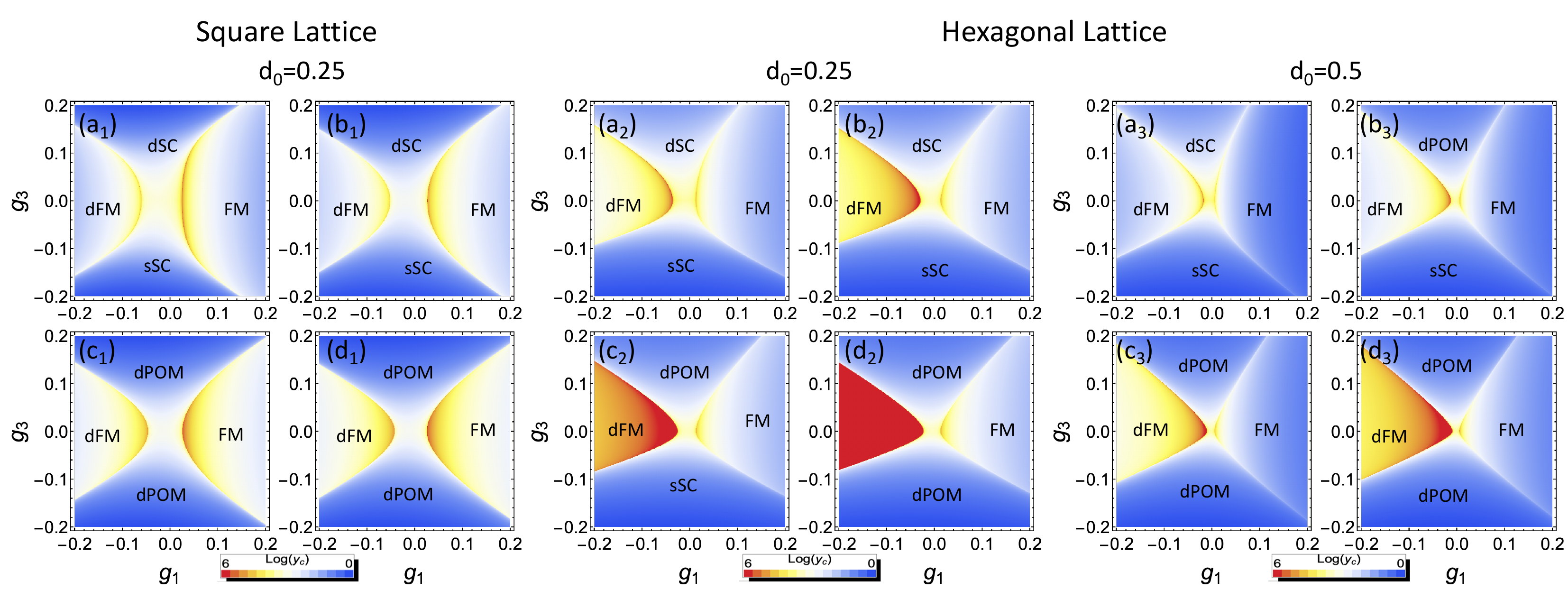}
\caption{The phase diagrams for the square and hexagonal lattices in the weak coupling space $(g_1(0),g_3(0))$ with fixed initial repulsive intra- and inter-patch forward coupling strength $g_2(0)=g_4(0)=0.1$. Here subgraphs (a$_{1,2,3}$), (b$_{1,2,3}$), (c$_{1,2,3}$) and (d$_{1,2,3}$) represent different fermion flavors $N_f=2$, $4$, $6$ and $8$, respectively. The left four panels display the case of the square lattice with $N_p =2$ and $d_0 = 0.25$, while the right eight panels show the case of the hexagonal lattices with $N_p = 3$ and $d_0 = 0.25$ and $d_0 = 0.5$. In all panels there are four different phases, which are either superconducting (SC), 
ferromagnetic (FM), or Pomeranchuk (POM) with s-wave (s) or d-wave (d) form factor. The color in the phase diagrams encodes the information of critical RG time $y_c$ as $\ln(y_c)$.}
\label{fig:Phasediagrams}
\end{figure*}
Driven by the recent progresses in the experimental and theoretical studies in the Kagome material AV$_3$Sb$_5$ with the hexagonal structure~\cite{Ortiz2019,Ortiz2020,Neupert,KJiang,Kang2022,Hu2021}, we first solve the RG equations (\ref{RGeq1})-(\ref{RGeq4}) for the hexagonal lattice with patches $N_p=3$ and different $N_f$ (the different orbital degeneracies). As shown in Fig. \ref{fig:RG flows}, we plot the representative RG flows of coupling strength $g_i(y)$ and
reduced parameters $\gamma_i(y)$ (insets) under the RG time $y$ with initial repulsive interactions $g_2(0)=g_3(0)=g_4(0)=0.1$ and $g_1(0)=0.01$, and a reasonable nesting parameter $d_0=0.5$. The three plots are for the number of flavors $N_f=2$, $4$ and $6$, respectively. For the repulsive coupling strength region $g_i(0)=g_0$ ($i=2,3,4$) and $g_1(0)\ll g_0$, $g_1(y)$ always tends to vanish, while $g_4(y)$ eventually changes sign and diverges to negative infinity. Thus, in this region, the negative $G_4$ will favor the charge Pomeranchuk and superconducting states. The detailed numerical results shown in Fig. \ref{fig:RG flows} indicate that $g_2(y)$ and $g_3(y)$ flow to positive infinite, while it is subtle to tell which one diverges first. For the normal case with $N_f=2$, Fig. \ref{fig:RG flows} (a) and (a$_1$) show that $g_3(y)$ behaves as $g_3(y) \sim 1/(y_c-y)$, i.e., with a similar power-law divergence as $g_4 (y)$, while $g_2(y)$ has a much slower power-law divergence  $1/(y_c -y )^\alpha$ with the exponent $0<\alpha\approx -2d_0(N_f-1) G_4<1$. In this case, the fixed trajectory is $(0,0,G_3,G_4)$ with $G_3>0$ and $G_4<0$. For the case with $N_f=4$ (from the orbital degeneracy), the flow of interactions are depicted in Fig. \ref{fig:RG flows} (b) and (b$_1$), both $g_2(y)$ and $g_3(y)$ will diverge as $g_{2/3}(y)\sim 1/(y_c-y)$. As a consequence, the fixed trajectory for this case is $(0,G_2,G_3,G_4)$ with $G_2, G_3>0$ and $G_4<0$. For the case with $N_f=6$, as shown in Fig. \ref{fig:RG flows} (c) and (c$_1$), $g_2(y)$ and $g_3(y)$ have a reversed behavior compared to the case of $N_f=2$, $g_2(y)$ scales as $1/(y_c-y)$, but $g_3(y)$ behaves slower as $1/(y_c-y)^{\alpha^{\prime}}$ with $0<\alpha^{\prime}\approx -2G_4<1$. Here the fixed trajectory is $(0,G_2,0,G_4)$ with $G_2>0$ and $G_4<0$.

To clearly demonstrate the evolution of relative interactions, we plot $\Gamma_i$ as function of the number of flavors $N_f$ in Fig. \ref{fig:GammaOrder} (a). $|\Gamma_2|$ becomes nonzero from $N_f=2.2$ and increases with increasing $N_f$. In contrast, $|\Gamma_3|$ decreases with increasing $N_f$ and drops to zero at $N_f=4.2$. These behaviors are intimately related to the RG equations. From the Eq. (\ref{RGeq4}), we find that $g_4$ diverges faster and the critical $y_c$ becomes smaller with increasing $N_f$, which is already numerically verified in Fig. \ref{fig:RG flows}. The more divergent negative $g_4$ and increasing $N_f$ have a positive feedback for the $g_2$ from the Eq. (\ref{RGeq2}), significantly enhancing $g_2$. While, as the one-loop correction of $g_3$ is decoupled from $g_2$ and does not depend on $N_f$, $g_3$ is slightly enhanced. The reduction of critical RG time $y_c$ with increasing $N_f$ will lead to a decreasing $\Gamma_3$. The relative evolution of $G_2$ and $G_3$ in the regions $g_0 > 0$ and $g_0 \gg g_1 (0) > 0$ with increasing flavor number $N_f$ for the hexagonal lattice will generate a phase transition from a particle-particle instability to a particle-hole instability. Here, with increasing $N_f$, $\lambda_{dSC}$ decreases but $\lambda_{dPOM}$ increases monotonically, as shown in Fig. \ref{fig:GammaOrder} (b), and thus the phase transition is from the $d$-wave superconducting state to the $d$-wave Pomenrunck state. The phase transition occurs around $N_f=3$. With an initial attractive $g_3$, the behaviors are similar to the case of repulsive $g_3$ as shown in Fig. \ref{fig:GammaOrder} (c), where $|\Gamma_2|$ increases and $|\Gamma_3|$ decreases with increasing $N_f$. The difference is that the phase transition is from the $s$-wave superconducting state to the $d$-wave Pomenrunck state and occurs at a larger $N_f=4.2$ as illustrated in Fig. \ref{fig:GammaOrder} (d).

We further study the phase diagrams of competing states in a wider range of initial interactions on the square and hexagonal lattice. In Fig. \ref{fig:Phasediagrams}, we take the initial intra- and inter-patch forward coupling strength as repulsive with $g_4(0)=g_2(0)=0.1$ and plot the diagrams in the coupling space $(g_1(0),g_3(0))$ for the square and hexagonal lattices. The small subplots (a$_{1,2,3}$), (b$_{1,2,3}$), (c$_{1,2,3}$) and (d$_{1,2,3}$) are calculated with the number of fermion flavors being $N_f=2$, $4$, $6$ and $8$, respectively. The two lattice structures share very similar phase diagrams for the same initial parameters, distinct from the case of conventional VHSs. Furthermore, one can observe that phase diagrams are separated into four regions. The ferromagnetic and anti-ferromagnetic (d-wave ferromagnetic) orders are robust in a wide range of coupling space for different nesting parameter $d_0$ and fermion flavors. For large initial negative and positive $g_3$, the system with small $N_f$ favors an s-wave and d-wave superconducting state, respectively. With the gradual increase of $N_f$, these orders turn to be the dPOM state. The transition from superconducting states to dPOM occurs at smaller $N_f$ with a larger nesting parameter $d_0$, as shown in phase diagrams (Fig. \ref{fig:Phasediagrams} (b$_2$) and (b$_3$)) of the hexagonal lattice with $d_0=0.25$ and $d_0=0.5$. In the square lattice with $N_p=2$, the phase diagrams with $d_0=0.5$ are similar to the case with $d_0=0.25$ and they are displayed in the Appendix \ref{sec:phasesquare}. A supermetal phase, exhibiting power-law divergent DOS but no long-range order~\cite{Isobe2019}, can appear on the hexagonal lattice at $d_0=0.25$ and $N_f=8$ (red regime of the sub-figure (d$_2$) of Fig. \ref{fig:Phasediagrams}). The phase diagrams with a variation of other couplings $g_2(0)$ and $g_4(0)$ are provides in the appendix \ref{sec:phasenegative}. They also display a similar feature that the instabilities in the particle-particle channel are suppressed but particle-hole orders are favored with increasing flavor number $N_f$.

\section{Discussion}
\begin{figure}[ht]
\centerline{\includegraphics[width=1\columnwidth]{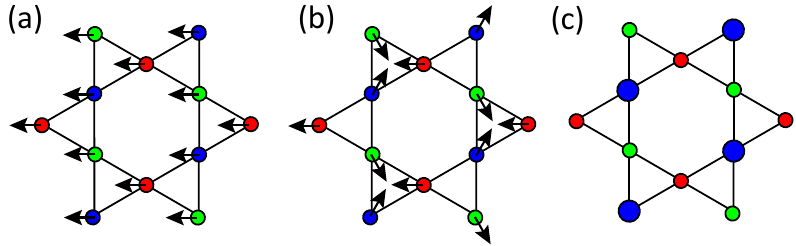}}
\caption{ Competing states in the kagome lattice: (a) ferromagnetic order, (b) 120$^{\circ}$ antiferromagnetic order (c) nematic order. The size of a circle represents the onsite energy of the corresponding sublattice.
 \label{realspace} }
\end{figure}

We discuss the ground state for the dominant states with degeneracy. On the square lattice, there are two saddle points and $N_p=2$. All competing states are non-degenerate, consistent with lattice symmetry. In the hexagonal lattice, such as, honeycomb or kagome lattices, there are three saddle points and $N_p=3$. There is a single $s$-wave state and two-fold degenerate $d$-wave states in each channel of instability, including superconductivity, spin density wave and charge density wave, with zero momentum transfer (Pomeranchuk order). In the following, we take the kagome lattice as an example. In the superconducting channel, for the $d$-wave pairing state, the system tends to form a $d+id$ state\cite{Chubukov2012,Classen2020} in order to maximize the condensation energy. This state is time-reversal breaking and carries an even Chern number. In the channel of spin density wave, the $s$-wave state is the ferromagnetic state and the magnetic moments of the three patches are aligned along the same direction, as shown in Fig. \ref{realspace} (a). The $d$-wave states are two-fold degenerate and the corresponding eigenvectors are $\phi^d_{1}=\frac{1}{\sqrt{6}}(1,1,-2)$ and $\phi^d_{2}=\frac{1}{\sqrt{2}}(1,-1,0)$. The order parameter can be written as $\bm{S}_{d}=\bm{s}_1\phi^d_1+\bm{s}_2 \phi^d_2$ and the corresponding free energy up to quartic order in $\bm{s}_{1,2}$ reads\cite{Classen2020},
\begin{eqnarray}
F^M_{d}=&\frac{\alpha}{2}(\bm{s}^2_1+\bm{s}^2_2)+\frac{\beta}{2} [(\bm{s}^2_1+\bm{s}^2_2)^2\nonumber \\
&-\frac{4}{3}\bm{s}^2_1\bm{s}^2_2+\frac{4}{3}(\bm{s}_1\cdot\bm{s}_2)^2].
\end{eqnarray}

As $\beta$ is generally greater than zero, by minimizing the free energy, we obtain that $\bm{s}_1$ and $\bm{s}_2$ are orthogonal with the same amplitude. Setting these two vectors to be the in-plane unit vectors $\bm{e}_x$ and $\bm{e}_y$, the corresponding order parameter is $\bm{S}_d=\frac{1}{\sqrt{6}}(\bm{e}_x+\sqrt{3}\bm{e}_y,\bm{e}_x-\sqrt{3}\bm{e}_y,-2\bm{e}_x)$. In the momentum space, the spin vector winds twice on the Fermi surface. For the $p$-type VHS in the kagome lattice, where each sublattice dominates the eigen state at each saddle point, this order corresponds to the intra-unit-cell antiferromagnetic order with an angle of 120$^\circ$ between nearby spins, as shown in Fig. \ref{realspace} (b). For the $m$-type VHS, where the eigenstate at each saddle point is attributed to a mixture of two sublattices, this order corresponds to both onsite and bond orders and thus there is no clear real-space pattern. In the channel of charge density wave, the $s$-wave state is isotropic and preserve all lattice symmetry. While, the $d$-wave states are two-fold degenerate and the eigen vectors are the same as the $d$-wave spin density wave state. The order parameter can be written as $\eta^C_{d}=\eta_1 \phi^d_{1}+\eta_2 \phi^d_{2}$. The free energy reads\cite{Kiesel2013,Hecker2018,Little2020,Classen2020},
\begin{eqnarray}
F^C_d=\frac{\alpha'}{2}(\eta^2_1+\eta^2_2)+\frac{\beta'}{2}(\eta^3_-+\eta^3_+),
\end{eqnarray}
where $\eta_{\pm}=\eta_1\pm i\eta_2$ and the cubic term is attributed to the six-fold rotation of the lattice. Assuming $\eta_{\pm}=\eta e^{\pm2i\theta}$ with $\theta$ being the orientation of the nematic director, the cubic term can be written as $\beta'\eta^3 cos(6\theta)$, which is minimized by $2\theta=2n\pi/3$ for $\beta'<0$ and $2\theta=(2n+1)\pi/3$ for $\beta'>0$. The corresponding order parameters are $\eta^C_{d}=\frac{1}{\sqrt{6}}(2,-1,-1)$, $\frac{1}{\sqrt{6}}(-1,2,-1)$ or $\frac{1}{\sqrt{6}}(-1,-1,2)$. These orders break the lattice rotational symmetry and thus are nematic. For the $p$-type VHS, this order corresponds to distinct onsite energies of sublattices (see Fig. \ref{realspace} (c)) or anisotropic intrasublattice hopping. For the m-type VHS, sublattices are mixed and this order is attributed to distinct onsite energies of sublattice and anisotropic intra- and inter-sublattice hopping.

Let us now discuss the implications of our findings for the kagome metals AV$_3$Sb$_5$. In AV$_3$Sb$_5$, both conventional and higher-order VHSs appear near the Fermi level~\cite{Kang2022,Hu2021}. The CDW order may originate from the electronic correlation effect enhanced by conventional VHSs~\cite{Feng2021,Denner2021,Lin2021,Park2021} or phonon softening~\cite{HXTan2021}. According to ARPES measurements, the momentum-dependent CDW-induced gaps are maximum at the M point and the conventional VHSs are shifted away from the Fermi level~\cite{Luo2022,YongHu2022,Kang2022}. In contrast, the higher-order VHS only exhibits slight energy shifting~\cite{YongHu2022} due to its poor Fermi surface nesting and is still close to the Fermi level in the CDW phase. Superconductivity and nematicity emerging inside the CDW order can derive from the instabilities of this higher-order VHS. The two-fold dPOM states will mix and form a nematic state breaking the lattice rotational symmetry (see Fig. \ref{realspace} (c)), which may account for the observed nematic order in transport measurements~\cite{HaiHuWen} and STM measurements~\cite{chenxianhui2022}. The superconductivity in experiments is $d$-wave or $s$-wave depending on the initial value of the interaction $g_3$ according to our calculations. With increasing pressure, the CDW and nematic orders get suppressed while superconductivity is enhanced as the both conventional and higher-order VHSs gradually move away from the Fermi level~\cite{LaBollita2021,Consiglio2022}. These are consistent with experimental observations.

\section{Conclusions and Remarks}\label{conclusion}
In summary, we systematically study the competing instabilities on the 2D square and hexagonal lattices at higher-order van Hove filling with an SU($N_f$) flavor degeneracy. These higher-order VHSs host the power-law divergent DOS and feature poor Fermi surface nesting. Firstly, we study the evolution of the band structure and explore the realization of higher-order VHSs on square and kagome lattices, based on tight-binding Hamiltonian and effective model around saddle points. Secondly, by performing renormalization group calculations, we investigate the competing orders with a variation of initial coupling constants and number of fermion flavors. Interestingly, we find that increasing flavor number can drive a generic phase transition from a superconductivity order to a nematic order. It originates from the significant enhancement of the inter patch density interaction $g_2$ with a larger flavor number $N_f$. Thirdly, the mean-field theory is used to determine the ground states and their real space patterns are also revealed on the kagome lattices. Implications for observed intriguing states in AV$_3$Sb$_5$ kagome metals are also discussed.

\section{Acknowledgments}
X. L. Han acknowledges supports from  China Postdoctoral Science Foundation Fellowship (No. 2022M723112).

\appendix
\section{Fixed trajectories and their stabilities on the square lattice}\label{FixedTSquare}
\begin{table*}\label{Tbl:Solutions}
	\centering
	\caption{Fixed trajectories and stable conditions on the square lattice}
	\label{table:Np2}
	\begin{tabular}{c|c|c|c|c|c|c|c}
	\hline
	\hline
No.	&$G_1$ & $G_2$ & $G_3$ & $G_4$ & conditions for fixed trajectories & leading order &stable condition \\
	\hline
	$1$&$0$ & $0$ & $0$ & $\frac{-1}{d_0(N_f-3)+1}$ & $N_f\neq 3-\frac{1}{d_0}$ & sPom/dPom & unstable  \\
	\hline
	$2$&$0$ & $\sqrt{\frac{d_0N_f+d_0-1}{4d_0^3N_f(N_f-1)^2}}$ & $0$ & $\frac{-1}{2d_0(N_f-1)}$ &$N_f\ge \frac{1}{d_0}-1$  &dPom  & $N_f\ge 1+\frac{1}{d_0}$ \\
	\hline
	$3$ & $0$ & $-\sqrt{\frac{d_0N_f+d_0-1}{4d_0^3N_f(N_f-1)^2}}$ & $0$ & $\frac{-1}{2d_0(N_f-1)}$ & $N_f\ge \frac{1}{d_0}-1$ &sPom & $N_f\ge 1+\frac{1}{d_0}$ \\
	\hline
	$4$ & $0$ & $0$ & $\frac{\sqrt{1-d_0(N_f-3)}}{2}$ & $-\frac{1}{2}$ & $N_f\le\frac{1}{d_0}+3$ & dSC or sPom/dPom & $N_f \le 3+\frac{1}{d_0}$\\
	\hline
	$5$ & $0$ & $0$ & $-\frac{\sqrt{1-d_0(N_f-3)}}{2}$ & $-\frac{1}{2}$ & $N_f\le\frac{1}{d_0}+3$ & sSC & $N_f \le 1+\frac{1}{d_0} $\\
	\hline
	$6$ & $N_f G_2$ & $\sqrt{\frac{1+d_0(N_f-1)}{4d_0^3(N_f+N_f^2)}}$ & $0$ & $\frac{1}{2d_0}$ & $N_f\ge 1$ &FM  & $N_f \ge 1$\\
	\hline
	$7$ & $N_f G_2$ & $-\sqrt{\frac{1+d_0(N_f-1)}{4d_0^3(N_f+N_f^2)}}$ & $0$ & $\frac{1}{2d_0}$ & $N_f\ge 1$ &dFM & $N_f\ge 1$\\
	\hline
	\hline
	\end{tabular}
\end{table*}
In this section, we can get the equations for the fixed trajectories by putting the relation $g_i(y)=G_i/(y_c-y)$ near critical time $y_c$ into Eq. (\ref{RGeq1})-(\ref{RGeq4}) as following,
\bea
&&G_1=2d_0G_1G_4, \\
&&G_3=-2G_3 G_4, \\
&&G_2=-2d_0G_4\big( (N_f-1)G_2-G_1 \big),\\
&&G_4=-G_4^2-G_3^2-d_0 N_fG_2^2\nonumber \\
&&-2d_0G_1G_2-d_0G_1^2-d_0G_4^2,
\eea
for the square lattice case with $N_p=2$. The fixed trajectories can be analyzed by using the stability matrix mentioned in the section. \ref{sec:FixedTandStb}. We summarize the solutions and their stability conditions in the table. \ref{Tbl:Solutions}.
\section{Fixed trajectories on the hexagonal lattice}\label{appendixB}
The equations of fixed trajectories on the hexagonal lattice with $N_p=3$ are given as,
\bea
&&G_1=d_0G_1\big(2G_4+G_1\big), \\
&&G_2=-d_0G_2\big( N_fG_2-2G_1 \big)\nonumber \\
&&-2d_0G_4\big((Nf-1)G_2-G_1 \big),\\
&&G_3=-G_3\big(G_3+2G_4\big), \\
&&G_4=-d_0\big(2N_f G_2^2-4G_1G_2-2G_1^2 \nonumber \\
&&+(N_f-3)G_4^2\big)-G_4^2-2G_3^2.
\eea
We solve these equations, and we list these solutions as below.

i: when $G_1=G_3=0$. $G_2$ and $G_4$ are given
\begin{eqnarray}
(1) \quad  &&G_2=0,G_4=\frac{-1}{1+d_0(N_f-3)},  \\
(2)\quad  &&G_4=\frac{8d_0-9N_f d_0 \pm S_2}{2(N_f d_0+8d_0^2-19N_f d_0^2+9N_f^2 d_0^2)} ,\nonumber \\
&& G_2= \frac{-N_f+2N_f d_0\mp (N_f-1)S_2}{N_f d_0[N_f+d_0(8-19N_f +9N_f^2)]},
\end{eqnarray}

ii: at the relation $G_1=\frac{1}{d_0}-2G_4, G_3=0$, we solve as
\begin{eqnarray}
(3)\quad  && G_2=-2G_4=\frac{17d_0\pm S_3}{d_0-27d_0^2+9N_f d_0^2}, \\
(4)\quad  &&G_2=\frac{N_f-2N_fd_0+N_f^2d_0\pm S_4}{N_f (N_f-8d_0-11N_f d_0+N_f^2 d_0)},\nonumber \\
&&G_4=\frac{-8d_0 -9N_f d_0\mp S_4}{2(N_fd_0-8d_0^2-11N_f d_0^2+N_f^2 d_0^2)},
\end{eqnarray}

iii: when $G_1=G_2=0$ and $G_3=-1-2G_4$, we have,
\bea
(5) \quad \quad \quad \quad \quad \quad G_4=-\frac{9\pm S_5}{2(9-3d_0+N_f d_0)},
\eea
iv: when $G_1=0$ and $G_3=-1-2G_4$, we can solve,
\bea
(6)\quad &&G_4=\frac{8d_0-17N_fd_0\mp S_6}{2(9N_fd_0+8d_0^2-19N_fd_0^2+9N_f^2 d_0^2)}, \nonumber \\
&&G_2=\frac{-9N_f d_0-6N_f d_0^2 +8 N_f^2 d_0^2\pm d_0(N_f-1)S_6}{9N_fd_0+8d_0^2-19N_fd_0^2+9N_f^2 d_0^2},\nonumber \\
\eea

v: when $G_1=\frac{1}{d_0}-2G_4$ and $G_3=-1-2G_4$, we can solve,
\bea
(7)\quad \quad &&G_2=\frac{25d_0\pm S_7}{9(d_0-3d_0^2+N_f d_0^2)},\nonumber \\
&&G_4=\frac{-25d_0 \mp S_7}{18(d_0-3d_0^2+N_f d_0^2)}, \\
(8)\quad \quad &&G_2=\frac{9N_fd_0+6N_fd_0^2+N_f^2d_0^2\pm d_0S_8}{N_f d_0(9N_f d_0-8d_0^2-11N_f d_0^2 +N_f^2 d_0^2)}, \nonumber \\
&&G_4=\frac{-8d_0-17N_f d_0\mp S_8}{2(9N_f d_0-8d_0^2-11N_f d_0^2 +N_f^2 d_0^2)},\nonumber \\
\eea
here we define the following formulas as,
\bea
&&S_{2}=\sqrt{N_f d_0 (-8+8d_0+9N_f d_0)},\\
&&S_3=\sqrt{d_0(8+73d_0+72N_f d_0)},\\
&&S_4=\sqrt{N_fd_0(8+8N_f-8d_0 +N_f d_0 +8N_f^2 d_0)},\\
&&S_5=\sqrt{9+24d_0-8 N_f d_0},\\
&&S_6=(-72N_f d_0-120N_f d_0^2+145N_f^2 d_0^2\nonumber \\
&&-64N_f d_0^3+152N_f^2d_0^3-72N_f^3 d_0^3)^{1/2}\\
&&S_7=\sqrt{d_0(72+337d_0+72N_fd_0+216d_0^2-72N_fd_0^2)},\nonumber \\ \\
&& S_8=\Big[(8d_0+17N_fd_0)^2-4(-2-2Nf+2N_f d_0)\nonumber \\
&&\times(9N_f d_0-8d_0^2-11N_f d_0^2+N_f^2 d_0^2) \Big]^{1/2}.
\eea
One can note the solutions of fixed trajectories on the hexagonal lattice are too complex to analysis their stable properties, in contrast to the case on the square lattice. 
\begin{figure}[ht]
\includegraphics[width=0.45\textwidth]{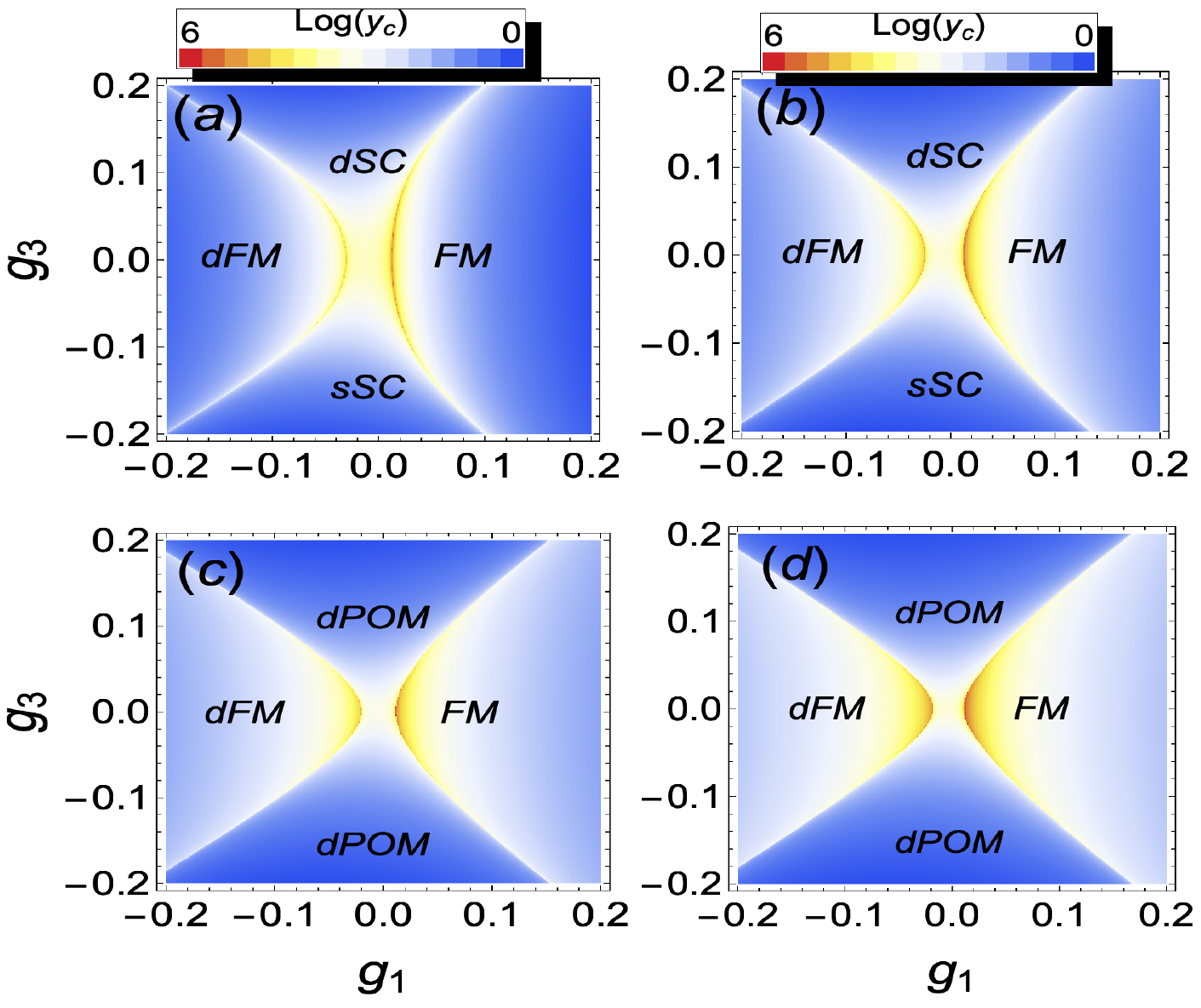}
\caption{ The phase diagrams for the square lattice in the weak coupling space $(g_1(0),g_3(0))$ with fixed initial repulsive intra- and inter-patch forward coupling strength $g_2(0)=g_4(0)=0.1$, and with the given nesting parameter $d_0=0.5$. Here (a-d) is for the case with $N_f=2$,$4$,$6$ and $8$, respectively.}
\label{fig:Np2d05}
\end{figure}

\section{Phase diagrams for the square lattice with the large nesting parameter}\label{sec:phasesquare}
The phase diagram with the nesting parameter $d_0=0.5$ in the square lattice is given in Fig. \ref{fig:Np2d05}. It is clear to see there is no dramatic difference comparing $d_0=0.25$ and $d_0=0.5$ in the given initial coupling constant $(g_2(0),g_4(0))=(0.1,0.1)$.

\section{Phase diagrams for the negative initial coupling constant}\label{sec:phasenegative}
\begin{figure*}[ht]
\includegraphics[width=0.98\textwidth]{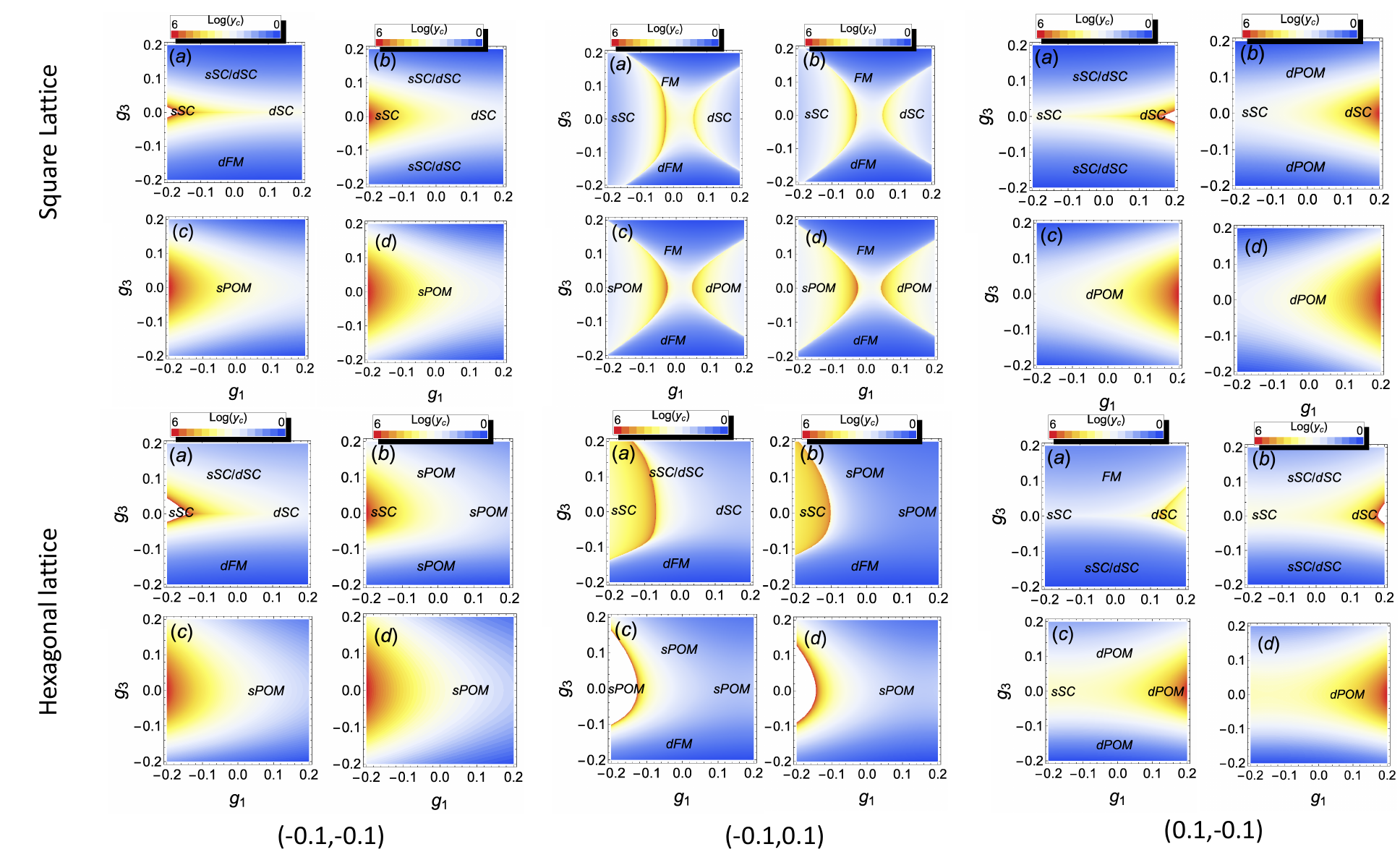}
\caption{ The phase diagrams for the square and hexagonal lattice in the weak coupling space $(g_1(0),g_3(0))$ with different fixed initial repulsive intra- and inter-patch forward coupling strength $(g_2(0),g_4(0))$, and with the given nesting parameter $d_0=0.25$. Here (a-d) is for the case with $N_f=2$,$4$,$6$ and $8$ respectively. The upper panel is for the square lattice, and down panel is for the hexagonal lattice. The initial given coupling constant $(g_2(0),g_4(0))$ is $(-0.1,-0.1)$ for the left panel, $(-0.1,0.1)$ for the middle panel, and $(0.1,-0.1)$ for the right panel respectively.}
\label{fig:phaseappendix}
\end{figure*}

We plot the phase diagrams with the nesting parameter $d_0=0.5$ in the hexagonal and square lattice for the different given initial coupling constant $(g_2(0),g_4(0))$. From the Fig. \ref{fig:phaseappendix}, one can note that there exist phase transition from superconductivity to the $s$- or $d$-wave charge order (sPOM or dPOM) as increasing the fermion flavor $N_f$.

\bibliographystyle{apsrev4-1}
\bibliography{KagomePRBv3.bib}

\end{document}